\documentclass[showpacs,twocolumn,aps,prl,floatfix,superscriptaddress,noshowpacs]{revtex4}
\usepackage{amsthm} 
\usepackage[mathlines]{lineno}

\usepackage{amsmath,amssymb,graphicx,bm,color,mathrsfs,verbatim,epstopdf,dcolumn,dsfont,cancel}

\usepackage{hyperref}
\hypersetup{ 
	colorlinks   = true
}

\definecolor{violet}{rgb}{0.58, 0.0, 0.83}
\newcommand{\mkadd}[1]{{\color{violet}{#1}}}

\newcommand{\gae}{\lower 2pt \hbox{$\,
\buildrel{\scriptstyle >}\over {\scriptstyle \sim}\,$}}
\newcommand{\lae}{\lower 2pt \hbox{$\,
\buildrel{\scriptstyle <}\over {\scriptstyle \sim}\,$}}

\begin{document}


\title{Schrieffer-Wolff Transformation for Periodically Driven Systems: Strongly Correlated Systems with Artificial Gauge Fields}%

\author{Marin Bukov}
\email{mbukov@bu.edu}
\affiliation{Department of Physics, Boston University, 590 Commonwealth Ave., Boston, MA 02215, USA}

\author{Michael Kolodrubetz}
\affiliation{Department of Physics, Boston University, 590 Commonwealth Ave., Boston, MA 02215, USA}
\affiliation{Department of Physics, University of California, Berkeley, CA 94720, USA}
\affiliation{Materials Sciences Division, Lawrence Berkeley National Laboratory, Berkeley, CA 94720, USA}

\author{Anatoli Polkovnikov}
\affiliation{Department of Physics, Boston University, 590 Commonwealth Ave., Boston, MA 02215, USA}

\date{\today}

\begin{abstract}
We generalize the Schrieffer-Wolff transformation to periodically driven systems using Floquet theory. The method is applied to the periodically driven, strongly interacting Fermi-Hubbard model, for which we identify two regimes resulting in different effective low-energy Hamiltonians. In the nonresonant regime, we realize an interacting spin model coupled to a static gauge field with a nonzero flux per plaquette. In the resonant regime, where the Hubbard interaction is a multiple of the driving frequency, we derive an effective Hamiltonian featuring doublon association and dissociation processes. The ground state of this Hamiltonian undergoes a phase transition between an ordered phase and a gapless Luttinger liquid phase. One can tune the system between different phases by changing the amplitude of the periodic drive.
\end{abstract}



\maketitle

The Schrieffer-Wolff transformation (SWT)~\cite{schrieffer_66,zhang_88,bravyi_11,barthel_09} is a generic procedure to derive effective low-energy Hamiltonians for strongly-correlated many-body systems. It allows one to eliminate high-energy degrees of freedom via a canonical transform. The SWT has proven useful for studying systems with a hugely degenerate ground-state manifold, such as the strongly-interacting limit of the Fermi-Hubbard model (FHM)~\cite{zhang_88}, without resorting to conventional perturbation theory.

Treating interactions in such a non-perturbative way is difficult in periodically-driven systems~\cite{dalessio_13,dalessio_14,lazarides_14,lazarides_14_2,ponte_15,bukov_15_erg}, which have received unprecedented attention following the realisation of dynamical localisation~\cite{dunlap_86,lignier_07,zenesini_09,creffield_10,gong_09}, artificial gauge fields~\cite{eckardt_10,struck_11,struck_13,aidelsburger_13,miyake_13,atala_14,kennedy_15}, models with topological~\cite{oka_09,kitagawa_11,grushin_14,jotzu_14,aidelsburger_14,flaeschner_15} and state-dependent~\cite{jotzu_15} bands, and spin-orbit coupling~\cite{galitski_13,jimenez-garcia_15}. In this paper, we consider strongly-interacting periodically-driven systems and show how the SWT can be extended to derive effective static Hamiltonians of non-equilibrium setups. The parameter space of such models, to which we add the driving amplitude and frequency, opens up the door to new regimes. We use this to propose realisations of nontrivial Hamiltonians, including  spin models in artificial gauge fields and the Fermi-Hubbard model with enhanced doublon association and dissociation processes.  

\begin{figure}
	\includegraphics[width=0.8\columnwidth]{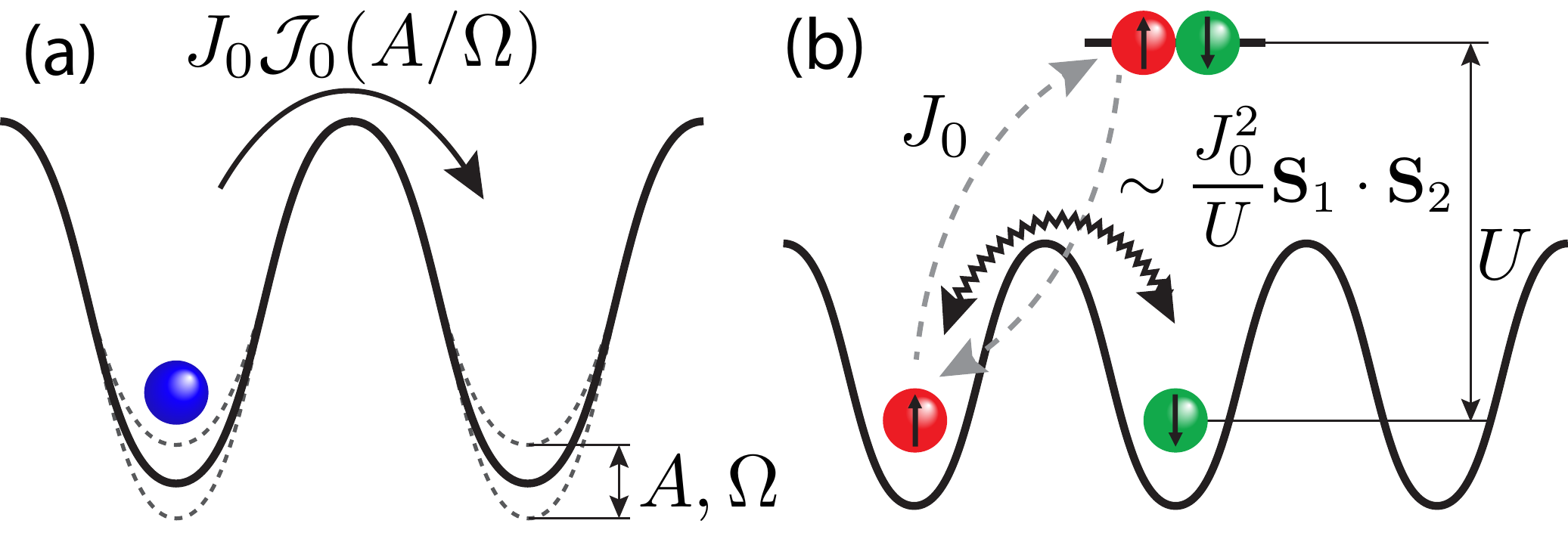}
	\caption{\label{fig:basic_process} Similarity between renormalisation of tunnelling, an interference effect induced virtually by an off-resonant drive (a), and Heisenberg interactions induced by virtual off-resonant interaction processes (b).}
\end{figure} 

\emph{SWT from the High-Frequency Expansion---}Intuitively, the high-frequency expansion for periodically-driven systems (HFE) and the SWT share the same underlying concept: they allow for the elimination of virtually-populated high-energy states to provide a dressed low-energy description, as illustrated in Fig.~\ref{fig:basic_process}. For a system driven off-resonantly (Fig.~\ref{fig:basic_process}a), virtual absorption of a photon renormalises tunnelling. Similarly, non-driven fermions develop Heisenberg interactions via off-resonant (virtual) tunnelling processes (Fig.~\ref{fig:basic_process}b). In this paper we combine the HFE and SWT into a single framework allowing one to treat both resonantly and non-resonantly driven systems on equal footing. Let us illustrate the connection by deriving the SWT using the HFE. Consider the non-driven FHM:
\begin{equation}
H = -J_0\sum_{\langle ij\rangle,\sigma}c^\dagger_{i\sigma}c_{j\sigma} + U\sum_j n_{j\uparrow}n_{j\downarrow},
\label{eq:FHM_nondriven}
\end{equation}
where $J_0$ is the bare hopping and $U$ is the fermion-fermion interaction. We are interested in the strongly-correlated regime $J_0\ll U$. Going to the rotating frame $|\psi^\text{rot}(t)\rangle = V^\dagger(t)|\psi(t)\rangle$ w.r.t.~the operator $V(t) = \exp\left(-i U t \sum_j n_{j\uparrow}n_{j\downarrow} \right)$ eliminates the energy $U$ in favor of fast oscillations. If $i \mathrm{d}_t |\psi^\text{rot}\rangle= H^\text{rot}(t) |\psi^\text{rot}\rangle$, then
\begin{eqnarray}
\! H^\text{rot}(t)&=&\!\!-\!J_0\!\sum_{\langle ij\rangle,\sigma}\!\left[g_{ij\sigma}\!+\!\left( e^{i Ut}h^\dagger_{ij\sigma} \!+\! \text{h.c.}\right)\right],\\
h^\dagger_{ij\sigma} &=& n_{i\bar\sigma}c^\dagger_{i\sigma}c_{j\sigma}(1 - n_{j\bar\sigma}),\nonumber\\
g_{ij\sigma} &=& (1 - n_{i\bar\sigma})c^\dagger_{i\sigma}c_{j\sigma}(1 - n_{j\bar\sigma}) +  n_{i\bar\sigma}c^\dagger_{i\sigma}c_{j\sigma}n_{j\bar\sigma},\nonumber
\label{eq:eq:Hrot_HFM}
\end{eqnarray}
where $\bar\uparrow = \downarrow$ and vice-versa. The first term $g_{ij\sigma}$ models the hopping of doublons and holons, while the second term $h^\dagger_{ij\sigma}$ represents the creation and annihilation of doublon-holon pairs. Since $H^\text{rot}(t)$ is time-periodic with frequency $U$, we can apply Floquet's theorem~\cite{floquet_83}. 
Thus, the evolution of the system at integer multiples of the driving period $T_U = 2\pi/U$ [i.e.~stroboscopically] is governed by the effective Floquet Hamiltonian $H_\mathrm{eff}$. If we write $H^\text{rot}(t)=\sum_\ell H^\text{rot}_\ell e^{i \ell U t}$, the HFE gives an operator expansion for $H_\mathrm{eff}=H^\text{rot}_0 + \sum_{\ell > 0} [H^\text{rot}_{\ell}, H^\text{rot}_{-\ell}] / \ell U + O(U^{-2})$~\cite{rahav_03_pra,goldman_14,goldman_14_res,eckardt_15,itin_15,mikami_15}. The zeroth-order term $H_\mathrm{eff}^{(0)} = H^\text{rot}_0$ is the period-averaged Hamiltonian [here the doublon-holon hopping $g$], while the first-order term is proportional to the commutator $H^{(1)}_\mathrm{eff}\sim J_0^2[h^\dagger,h]/U$, cf.~Fig.~\ref{fig:basic_process}b:

\begin{equation}
H_\text{eff} \approx -J_0\sum_{\langle ij \rangle,\sigma}g_{ij\sigma} + \frac{4J_0^2}{U}\sum_{\langle ij \rangle} \left( {\bf S}_{i}\cdot {\bf S}_{j} - \frac{n_in_j}{4} \right).
\label{eq:XXZ_FHM}
\end{equation}
This effective Hamiltonian is in precise agreement with the one from the standard SWT~\footnote{The approximate sign in Eq.~\eqref{eq:XXZ_FHM} is used since we neglected part of the correction terms, cf.~Ref.~\cite{keeling_notes}.}. At half-filling, doublons and holons are suppressed in the ground state and this reduces to the Heisenberg model. Away from half-filling this Hamiltonian reduces to the $t-J$ model~\cite{zhang_88,keeling_notes}.

Using the HFE to perform the SWT offers a few advantages: (i) the SW generator comes naturally out of the calculation, (ii) one can systematically compute higher-order corrections~\cite{rahav_03_pra,goldman_14,bukov_14,goldman_14_res,eckardt_15,itin_15,mikami_15}, and (iii) the HFE allows for obtaining not only the effective Hamiltonian but also the kick operator, which keeps track of the mixing between orbitals and describes the intra-period dynamics~\cite{goldman_14, bukov_14}. This is important for identifying the fast timescale associated with the large frequency $U$ in dynamical measurements~\cite{trotzky_08}, and expressing observables through creation and annihilation operators dressed by orbital mixing~\cite{bukov_14}. 

\emph{Generalisation to Periodically-Driven Systems.---}The HFE allows us to extend the SWT to time-periodic Hamiltonians. Related approaches have been used to study non-interacting Floquet topological insulators~\cite{nakagawa_14} and ultrafast dynamical control of the spin exchange coupling~\cite{mentink_15} in fermionic Mott insulators~\cite{bermudez_15}. Let us add to the FHM an external periodic drive:
\begin{eqnarray}
H(t) \!=\! \!-\!J_0\!\sum_{\langle ij\rangle,\sigma}\!c^\dagger_{i\sigma}c_{j\sigma} \!+\! U\!\sum_j n_{j\uparrow}n_{j\downarrow}\!+\!\sum_{j\sigma}\! f_{j\sigma}(t)n_{j\sigma}.
\label{eq:FHM_driven}
\end{eqnarray}
The driving protocol $f_{j\sigma}(t)$ with frequency $\Omega$ encompasses experimental tools such as mechanical shaking, external electromagnetic fields, and time-periodic chemical potentials, relevant for the recent realisations of novel Floquet Hamiltonians. In the following, we work in the limit $J_0\ll U,\Omega$ and assume that the amplitude of the periodic modulation also scales with $\Omega$~\cite{bukov_14}.

Since both the interaction strength $U$ and the driving amplitude are large, we go to the rotating frame w.r.t.~$V(t) = e^{-i\left[ U t \sum_j n_{j\uparrow}n_{j\downarrow} + \sum_{j,\sigma} F_{j\sigma}(t)n_{j\sigma}\right] }$, where $F_{j\sigma}(t) = \int^t f_{j\sigma}(t')\mathrm{d}t'$. The drive induces phase shifts to the hopping:
\begin{eqnarray}
\! H^\text{rot}\!(t)\!=\!\!-\!J_0\!\sum_{\langle ij\rangle,\sigma}\! \left[ e^{i\delta F_{ij\sigma}(t) }g_{ij\sigma}\!+\!\left(\!e^{i\left[\delta F_{ij,\sigma}(t) \!+\! Ut\right]}h^\dagger_{ij\sigma} \!+\! \text{h.c.}\!\right) \right] \!\nonumber
\label{eq:Hrot_driven}
\end{eqnarray}
where $\delta F_{ij,\sigma}(t) =  F_{i\sigma}(t) - F_{j\sigma}(t)$. Notice that now there are two frequencies in the problem: $U$ and $\Omega$. Hence, $H^\text{rot}(t)$ is not strictly periodic in either. To circumvent this difficulty, we choose a common frequency $\Omega_0$ by writing $\Omega = k\Omega_0$ and $U = l\Omega_0$ where $k$ and $l$ are co-prime integers. Then $H^\text{rot}(t)$ becomes periodic with period $T_{\Omega_0} = 2\pi/\Omega_0$, and we can proceed using the HFE. Alternatively, before going to the rotating frame, we could decompose the interaction strength as $U = l\Omega + \delta U$,  where $\delta U$ acts as a detuning, and can continue without including the term proportional to $\delta U$ in $V(t)$.

\emph{Non-resonant Driving.---} Let us first assume $k, l \gg 1$ such that resonance effects can be ignored. We begin by Fourier-expanding the drive $e^{i \delta F_{i j \sigma}(t)} = \sum_\ell A^{(\ell)}_{i j \sigma} e^{i \ell \Omega t}$. If opposite spin species are driven out-of-phase, we have $A^{(\ell)}_{i j \bar \sigma} = (A^{(-\ell)}_{i j \sigma})^\ast$. Similarly, flipping the direction of the bond flips the sign of $\delta F$, so $A^{(\ell)}_{j i \sigma} = (A^{(-\ell)}_{i j \sigma})^\ast$. We now apply the generalised SWT with frequency $\Omega_0$. At half-filling and for off-resonant driving double occupancies are suppressed, and the dominant term in the effective Hamiltonian is $H_\mathrm{eff}^{(1)}=\sum_{\ell > 0} [H^\text{rot}_{\ell}, H^\text{rot}_{-\ell}] / \ell \Omega_0$. Two types of commutators occur in this sum: the first comes from terms that have no oscillation with frequency $U$, giving commutators of the form: $\left[ \sum_{ij\sigma} A^{(\ell)}_{ij\sigma} g_{i j \sigma}, \sum_{i'j'\sigma'} A^{(\ell)}_{i'j'\sigma'} g_{i' j' \sigma'}\right]$; all of these commutators vanish. The second type are the same commutators relevant for the SWT: $\left[ \sum_{ij\sigma} A^{(\ell)}_{ij\sigma} h^\dagger_{i j \sigma}, \sum_{i'j'\sigma'} A^{(-\ell)}_{i'j'\sigma'} h_{j' i' \sigma'} \right]$, but note the presence of all higher-order harmonics induced by the drive. These involve terms rotating with $e^{i (U + \ell \Omega) t}$, and thus will be suppressed by a $(U + \ell \Omega)$--denominator. The commutators are explicitly done in the Supplemental material~\cite{supplementary_SWT}, giving
\begin{equation}
H_\mathrm{eff}^{(1)} \!=\! \sum_{\langle ij\rangle,\ell} \frac{J_0^2}{U \!+\! \ell \Omega} \!\left(\! \alpha_{ij}^{(\ell)} S_i^+ S_j^- \!+\! \alpha_{ij}^{(\ell) \ast} S_i^- S_j^+ \!+\! 2 \beta_{ij}^{(\ell)} S_i^z S_j^z \!\right)\!,
\nonumber
\end{equation}
where $\alpha_{ij}^{(\ell)} \equiv A_{i j \uparrow}^{(\ell)} A_{i j \uparrow}^{(-\ell)}$ and $\beta_{ij}^{(\ell)} \equiv |A_{i j \uparrow}^{(\ell)}|^2$.

\begin{figure}
	\includegraphics[width=\columnwidth]{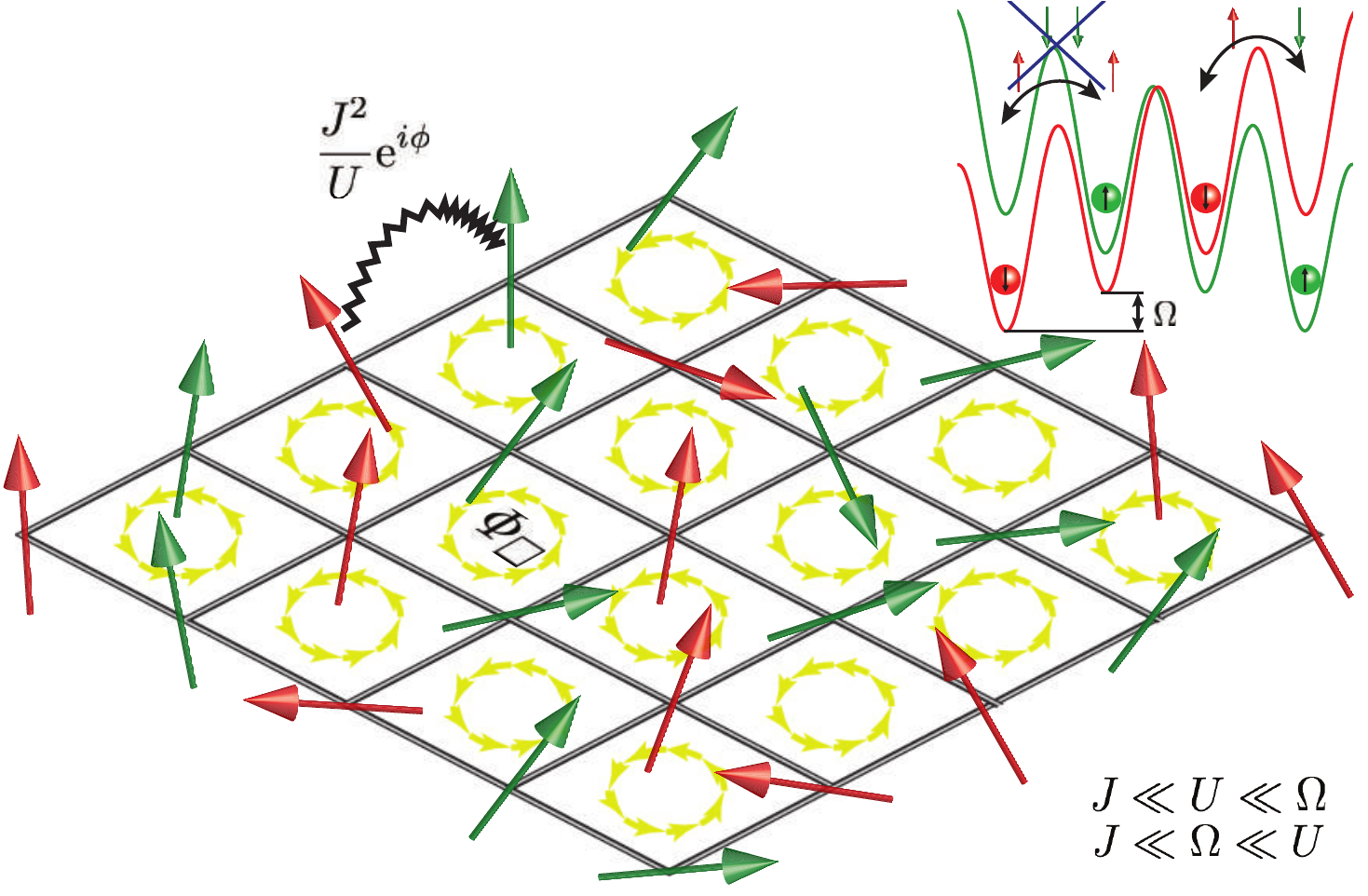}
	\caption{\label{fig:offresonant} In the presence of a spin-dependent drive off-resonant with the interaction strength $U$ (inset), the stroboscopic physics of the strongly-driven, strongly-correlated Fermi-Hubbard model is governed by an effective spin Hamiltonian in the presence of a gauge field.}
\end{figure}

One can Floquet-engineer the Heisenberg model with a uniform magnetic flux per plaquette $\Phi_\square$, see Fig.~\ref{fig:offresonant}. To this end, we choose the spin-dependent driving protocol $f_{j,\sigma}(t) = \sigma\left[A\cos\left(\Omega t + \phi_{j}\right) + \Omega m \right]$ (c.f.~Fig.~\ref{fig:offresonant}, inset),
where $\phi_j = \phi_{mn} = \Phi_\square(m+n)$, $\sigma \in \{ \uparrow,\downarrow \} \equiv \{1,-1\}$, and we denote the square-lattice position by ${\bf r}_j = (m,n)$. Such spin-sensitive drives are realised in experiments via the Zeeman effect using a periodically-modulated~\cite{jotzu_15} and static~\cite{aidelsburger_13,miyake_13} magnetic-field gradients which couple to atomic hyperfine states. For this protocol,
\begin{eqnarray*}
A_{(m,n),(m,n+1)\uparrow}^{(\ell)} \equiv A_{y\uparrow}^{(\ell)} &=& e^{i \ell \phi_{mn}} \mathcal{J}_\ell (2 \zeta_\Phi) \\
A_{(m,n),(m+1,n)\uparrow}^{(\ell)} \equiv A_{x\uparrow}^{(\ell)} &=& e^{i (\ell+1) \phi_{mn}} \mathcal{J}_{\ell+1} (2 \zeta_\Phi) ~,
\end{eqnarray*}
where $\mathcal{J}_\ell$ is the Bessel function of the first kind, $\zeta = A/\Omega$ is the dimensionless driving strength, and $\zeta_\Phi = \zeta \sin(\Phi_\square / 2)$ is the flux-modified strength~\footnote{There is an overall phase factor $(\Phi_\square + \pi)/2$ in $A_{x/y \uparrow}$ which is irrelevant to the global physics, so we gauge it away by rotating $S^+_{m n} \to S^+_{mn} e^{i m (\Phi_\square + \pi)}$, see Suppl.~\cite{supplementary_SWT}.}.

There are two physically interesting limits. For $U \ll \Omega$ only $\ell=0$ survives and we get
\begin{widetext}
\begin{equation}
\nonumber
	H_\text{eff}^{U\ll\Omega}  = \sum_{m,n } \Bigg(J^\text{ex,x}_\text{eff}\left[  S^z_{m+1,n}S^z_{mn} + \frac{1}{2}\left(e^{2i\phi_{mn} }S^+_{m+1,n}S^-_{mn} + \text{h.c.}\right)\right] +  J^\text{ex,y}_\text{eff}\left[  S^z_{m,n+1}S^z_{mn} + \frac{1}{2}\left(S^+_{m,n+1}S^-_{mn} + \text{h.c.}\right)\right]\Bigg), 
	\label{eq:Heis_gauge}
\end{equation}
where $J^\text{ex,x/y}_\text{eff} = 4\left[J\mkadd{_0}\mathcal{J}_{1/0}\left(2 \zeta_\Phi \right)\right]^2/U$. For $\Omega \ll U$, we can set $U + l \Omega \to U$ and sum over $l$ to obtain
\begin{equation}
\nonumber
	H_\text{eff}^{\Omega\ll U} = \frac{4J_0^2}{U}\sum_{m,n } \Bigg[  S^z_{m+1,n}S^z_{mn} + \frac{\mathcal{J}_2(4 \zeta_\Phi)}{2}\left(e^{2i\phi_{mn} }S^+_{m+1,n}S^-_{mn} + \text{h.c.}\right) +  S^z_{m,n+1}S^z_{mn} + \frac{\mathcal{J}_0(4 \zeta_\Phi)}{2}\left(S^+_{m,n+1}S^-_{mn} + \text{h.c.}\right)\Bigg]~.
\end{equation}
\end{widetext}
The exchange strengths depend on $\Omega$ and $U$, but both limits give spin Hamiltonians with phases along $x$. This phase physically appears on the flip-flop and not the Ising term because the drive is spin-dependent. Thus a phase difference only occurs if the electron virtually hops as one spin and returns as the other.

Let us discuss the regime $J_0\ll\Omega\ll U$ a bit more. This spin Hamiltonian can be identified with the Heisenberg model in the presence of an artificial gauge field with flux $\Phi_\square$ per plaquette. Whenever the $S^zS^z$-interaction is small, the Hamiltonian reduces to the fully-frustrated XY model in 2D, in which one cannot choose a spin configuration minimizing the spin-exchange energy for all XY-couplings. In the classical limit, similarly to a type-II superconductor, the minimal energy configuration is known to be the Abrikosov vortex lattice~\cite{teitel_83,ryu_97}. The realisation of the deep XY-regime with this particular driving protocol is limited, since $|\mathcal{J}_2(4 \zeta_\Phi)|< 1$ but, at finite $S^zS^z$--interaction a semi-classical study showed that vortices persist and can be thought of as half-skyrmion configurations of the Ne\'el field~\cite{lindner_09,lindner_10,wu_04}. Another interesting feature of the spin Hamiltonian is that it exhibits a Dzyaloshinskii-Moriya (DM) interaction term~\cite{cai_12,radic_12,cole_12,piraud_14}, ${\bf D}_{mn}\cdot \left( {\bf S}_{m+1,n}\times {\bf S}_{mn}\right)$. The DM coupling is spatially-dependent, polarised along the $z$-direction ${\bf D}_{mn} = \sin(\phi_{mn})\mathcal{J}_2(4 \zeta_\Phi )\hat{\bf n}_z/2$, and present only along the $x$-lattice direction. 

Finally, let us mention that spin-$1/2$ systems are equivalent to hard-core bosons. In this respect, $H_\text{eff}^{U\ll\Omega}$ and $H_\text{eff}^{\Omega\ll U}$ model hard-core bosons with strong nearest-neighbour interactions in the presence of a gauge field. For a flux of $\Phi_\square = \pi/2$ the non-interacting model has four topological Hofstadter bands. If we then consider the strongly-interacting model, and half-fill the lowest Hofstadter band ($S^z_\mathrm{tot} = -3 N_\mathrm{site}/8$), the Heisenberg model supports a fractional quantum Hall ground state~\cite{wang_11,regnault_11,hafezi_07,grushin_14}. Away from half-filling of the fermions, doublon and holon hopping terms appear in the effective Hamiltonian, cf.~Suppl.~\cite{supplementary_SWT} and it would be interesting to study the effect of such correlated hopping terms~\cite{kourtis_15} on this topological phase. 

\emph{Resonant Driving.---}Novel physics arises in the resonant-driving regime $J_0\ll U =l\Omega$. To illustrate this, we choose a one-dimensional system with the driving protocol $f_{j\sigma}(t) = jA\cos\Omega t$, which was realised experimentally by mechanical shaking~\cite{lignier_07,lignier_07,zenesini_09,creffield_10}. Unlike off-resonant driving, resonance drastically alters the effective Hamiltonian by enabling the lowest-order term $H^\text{{(0)}}_\text{eff}$: on resonance, the doublon-holon (dh) creation/annihilation terms $h^\dagger$ survive the time-averaging, and the leading-order effective Hamiltonian reads
\begin{eqnarray}	
\label{eq:Heff_resonant}
H_\text{eff}^{(0)} =  \sum_{\langle ij\rangle,\sigma} \left[  - J_\text{eff} g_{ij\sigma}  - K_\text{eff}\!\left( (-1)^{l\eta_{ij}} h^\dagger_{ij\sigma} + \text{h.c.}\right)\right],
\end{eqnarray}
where $\eta_{ij} = 1$ for $i>j$, $\eta_{ij} = 0$ for $i<j$, $J_\text{eff} = J_0\mathcal{J}_0(\zeta)$, and $K_\text{eff} = J_0\mathcal{J}_l(\zeta)$. The first term, $g_{ij\sigma}$, is familiar from the static SWT, with a renormalised coefficient $J_\text{eff}$. The term proportional to $h^\dagger_{ij\sigma}$ appears only in the presence of the resonant periodic drive and is the source of new physics in this regime. By adjusting the drive strength, one can tune $J_\text{eff}$ and $K_\text{eff}$ to a range of values, including zeroing out either one. Starting from a state with unpaired spins, dh pairs are created via resonant absorption of drive photons. Hence, holons and doublons become dynamical degrees of freedom governed by $H^\text{{(0)}}_\text{eff}$, with the Heisenberg model as a subleading correction. The dh production rates and further properties of the system have been investigated both experimentally and theoretically~\cite{kollath_06,huber_09,tokuno_11,tokuno_12,greif_11,sensarma_09,strohmaier_10,hassler_09,balzer_14,mentink_15,werner_15,bello_15}. A DMFT study found that the AC field can flip the band structure, switching the interaction from attractive to repulsive~\cite{tsuji_11}.  

\begin{figure}
	\includegraphics[width=\columnwidth]{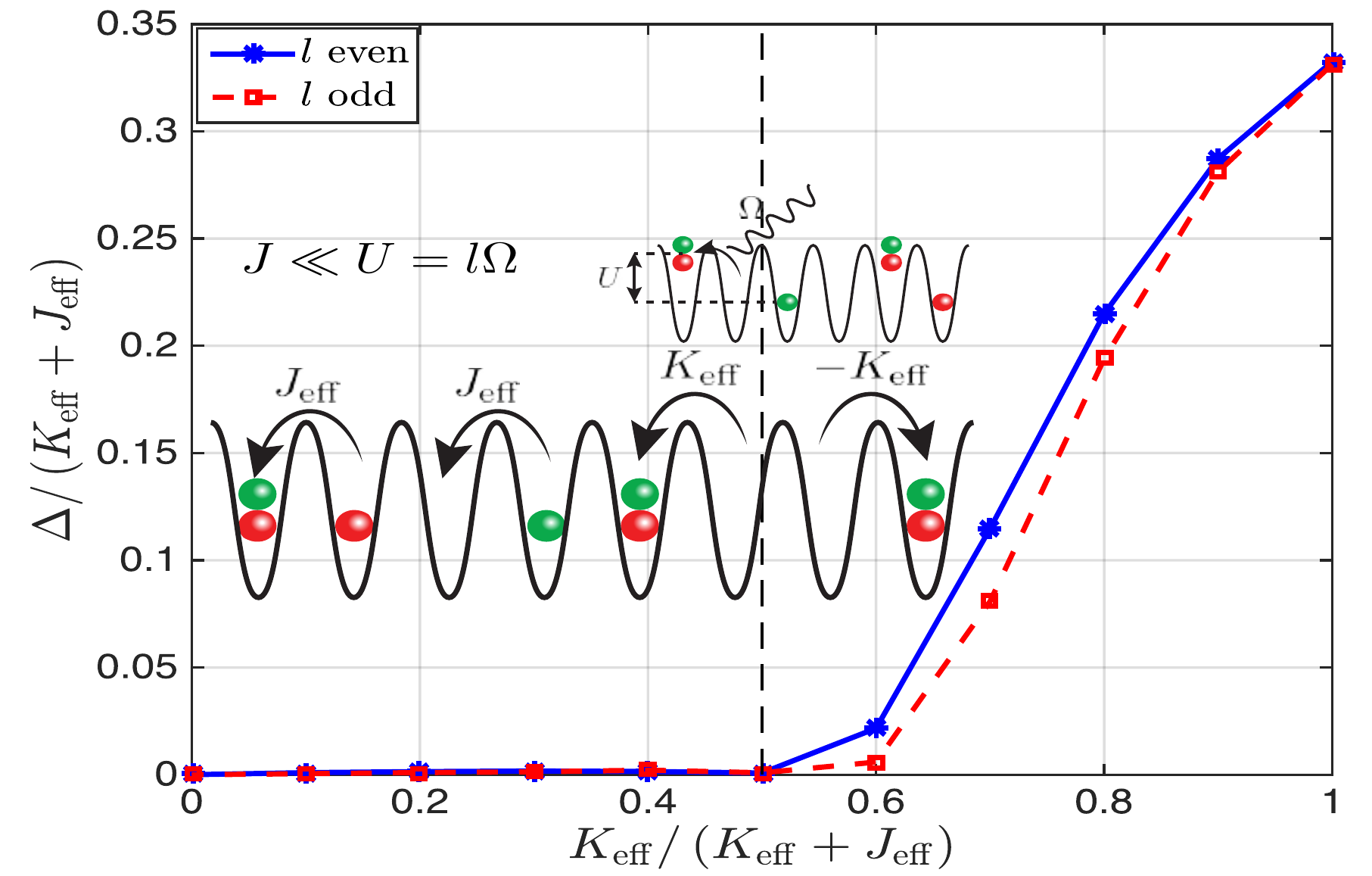}
	\caption{\label{fig:resonant} Resonant driving of the Fermi-Hubbard model enables doublon creation and dissociation processes (inset). The many-body gap $\Delta$ shows a phase transition from a gapless Luttinger liquid to gapped translation-invariance-broken phase. The doublon/holon hopping and creation coefficients $J_\mathrm{eff}$ and $K_\mathrm{eff}$ are controlled by varying the driving amplitude. }
\end{figure}

Such correlated hopping models have been proposed to study high-$T_c$ superconductivity~\cite{arrachea_94,aligia_00,aligia_99}. To get an intuition about the effect of the new terms, we use the ALPS DMRG and MPS tools~\cite{bauer_11,dolfi_14} to calculate the ground state of $H^\text{{(0)}}_\text{eff}$ at half-filling. The many-body gap in the thermodynamic limit $\Delta$ is extracted from simulations of even-length chains with open boundary conditions by extrapolation in the system size: $\Delta(L) = const/L + \Delta$. We numerically confirm that the model features a transition between a symmetry-broken ordered phase and a gapless Luttinger liquid phase~\cite{arrachea_94,aligia_00,aligia_99} as follows~\footnote{Periodically-driven Luttinger liquids were studied in Ref.~\cite{bukov_12}.}. For $K_\mathrm{eff} > J_\mathrm{eff}$, the physics is dominated by the dh creation/annihilation processes. In this regime, fermions can hop along the lattice by forming and destroying dh pairs. Thus, for \emph{$l$ even} the ground state exhibits bond-wave order with order parameter $B_j = \sum_{\sigma}c^\dagger_{j+1,\sigma}c_{j\sigma} + \text{h.c.}$, while the corresponding order parameter for $l$ odd is not yet known. This order breaks translation invariance with a 2-site unit cell, and thus yields a many-body gap for even-length chains with open boundary conditions (cf.~Fig.~\ref{fig:resonant}). For $K_\mathrm{eff} < J_\mathrm{eff}$, renormalization group arguments show that bond ordering terms become irrelevant, leading to a gapless Luttinger liquid~\cite{japaridze_99}. At $K_\mathrm{eff} = J_\mathrm{eff}$ and \emph{for $l$ even}, one  surprisingly finds that the system is equivalent to free fermions. The existence of such a non-interacting point is rather striking, since it means that a strongly-driven, strongly-interacting system can effectively behave as if the fermions were free. This phenomenon can be understood by noticing that double occupancies, effectively forbidden in the absence of the drive by strong interactions, are re-enabled by the resonant driving term. As a result, whenever the amplitude of the driving field matches a special value to give $K_\text{eff}=J_\text{eff}$, the matrix element for creation of doublons and holes becomes equal to their hopping rate and the effect of the strong interaction is completely compensated by the strong driving field. We emphasize that this is a highly non-perturbative effect since it requires a large drive amplitude $A\sim U = l\Omega$. 

It bears mention that all regimes of the model are accessible using present-day cold atoms experiments~\cite{greif_11}. We propose a loading sequence into the ground state of $H_\mathrm{eff}^{(0)}$ in the Supplemental material~\cite{supplementary_SWT}. Moreover, by tuning the frequency away from resonance, one can write $U = \delta U + l\Omega$ and go to the rotating frame w.r.t.~the $l\Omega$-term, keeping a finite on-site interaction $\delta U$ in the effective Hamiltonian. This is required if one wants to capture important photon-absorption avoided crossings in the exact Floquet spectrum. Including artificial gauge fields is also straightforward in higher dimensions, see Suppl.~\cite{supplementary_SWT} and expected to produce novel topological phases. By utilizing resonance phenomena, this scheme only requires shaking of the on-site potentials, which is easier in practice than other schemes which have suggested modulating the interaction strength to realize similar Hamiltonians~\cite{di_liberto_14,greschner_14}.


\emph{Discussion/Outlook.---}It becomes clear from the discussion above how to generalise the SWT to arbitrary strongly-interacting periodically-driven models: First, we identify the large energy scale denoted by $\lambda$ (e.g., $\lambda = U$) and write the Hamiltonian as $H = H_0 + \lambda H_1 +H_\mathrm{drive}(t)$. Second, we go to the rotating frame using the transformation $V(t)=\exp\left(-i\lambda t H_1 -i\int^t H_\mathrm{drive}(t')\mathrm{d}t' \right)$ to get a new time-dependent Hamiltonian with frequencies~\footnote{Formally, the identification of a well-defined frequency $\lambda$ in the rotating frame requires that the spectrum of $H_1$ is commensurate, which is the case whenever $H_1$ is a density-density interaction.} $\lambda$ and $\Omega$: $H^\text{rot}(t) = V^\dagger(t)H_0V(t)$.  Finally, depending on whether we want to discuss resonant or nonresonant coupling, we apply the HFE to obtain the effective Hamiltonian $H_\text{eff}$ order by order in $\lambda^{-1}$ and $\Omega^{-1}$. This procedure will generally work if a closed-form evaluation of $H^{\rm rot}(t)$ is feasible. For instance, $H_1$ can be a local Hamiltonian or can be written as a sum of local commuting terms. The method also works if the interaction strength is periodically modulated~\cite{di_liberto_14,greschner_14,wang_14}.

Although isolated interacting Floquet systems are generally expected to heat up to infinite temperature at infinite time~\cite{dalessio_13,dalessio_14,lazarides_14,lazarides_14_2,ponte_15,roy_15}, the physics of such systems at experimentally-relevant timescales is well-captured by the above effective Hamiltonians; indeed, it was recently argued that typical heating rates at high frequencies are suppressed exponentially~\cite{abanin_15,kuwahara_15,mori_15,abanin_15_2}, and long-lived pre-thermal Floquet steady states have been predicted~\cite{canovi_15,bukov_15_prl,kuwahara_15,abanin_15_2}. In particular, rigorous mathematical proofs~\cite{kuwahara_15,mori_15,abanin_15_2} supported by numerical studies~\cite{bukov_15_erg} showed that the mistake in the dynamics due to the approximative character of the HFE is under control for the large frequencies and the experimentally-relevant times considered. Our work paves the way for studying such strongly-driven, strongly-correlated systems. Both the resonant and non-resonant regimes that we analyse for the FHM yield systems directly relevant to the study of high-temperature superconductivity. More generally, we show that by using the generalised SWT, one can Floquet-engineer additional knobs controlling the model parameters of strongly-correlated systems, such as the spin-exchange coupling. Our methods are readily extensible to strongly-interacting bosonic systems, as well as many other systems under active research.

\begin{acknowledgments}
We thank L.~D'Alessio, E.~Altman, W.~Bakr, E.~Demler, M.~Eckstein, A.~Grushin, M.~Heyl, D.~Huse, A.~Iaizzi, G.~Jotzu, R.~Kaul, S.~Kourtis, M.Piraud, A.~Sandvik and R.~Singh for insightful and interesting discussions. We are especially grateful to M.~Dolfi and all contributors to the ALPS project~\cite{bauer_11,dolfi_14} for developing the ALPS MPS and DMRG tools used in this work. We thank A.~Rosch for pointing out to us the potential connection between the HFE and the SWT. This work was supported by AFOSR FA9550-13-1-0039, NSF DMR-1506340, and ARO  W911NF1410540. M.~K.~was supported by Laboratory Directed Research and Development (LDRD) funding from Berkeley Lab, provided by the Director, Office of Science, of the U.S. Department of Energy under Contract No. DE-AC02-05CH11231.
\end{acknowledgments}

\bibliographystyle{apsrev4-1}
\bibliography{Floquet_bib}

\begin{widetext}
 \newpage

\section{\large Supplemental Material}

\section{The High-Frequency Expansion.}
We open up the discussion by briefly reviewing the basic tool used in the main text -- the (van Vleck) High-Frequency Expansion (HFE). For a more-detailed description, consult Refs.~\cite{rahav_03_pra,goldman_14,bukov_14,goldman_14_res,eckardt_15,itin_15,mikami_15}. Consider a time-periodic Hamiltonian $H(t+T) = H(t)$. According to Floquet's theorem, the evolution operator $U(t_2,t_1)=\mathcal T\left[ \mathrm{exp}(-i\int_{t_1}^{t_2} H(t) \mathrm{d}t) \right]$, where $\mathcal T$ denotes time ordering, can be cast in the form
\begin{equation}
U(t_2,t_1) = \exp\left[-iK_\text{eff}(t_2)\right]\exp\left[-iH_\text{eff}(t_2-t_1)\right]\exp\left[iK_\text{eff}(t_1)\right],
\end{equation} 
with the time-independent effective Hamiltonian $H_\text{eff}$ governing the slow dynamics and the $T=2\pi/\Omega$-periodic kick operator $K_\text{eff}(t)$ describing the micromotion, i.e.~the fast dynamics within a period. In the high-frequency limit, one can calculate perturbatively the kick operator and the effective Hamiltonian as follows:
\begin{eqnarray}
H_\text{eff}^{(0)} &=& H_0 = \frac{1}{T}\int_0^T\mathrm{d}t\,H(t),\nonumber\\
H_\text{eff}^{(1)} &=& \frac{1}{\hbar\Omega}\sum_{\ell=1}^\infty \frac{1}{\ell} [H_\ell,H_{-\ell}] = \frac{1}{2!Ti\hbar}\int_{0}^{T}\mathrm{d}t_1\int_{0}^{t_1}\mathrm{d}t_2\, f(t_1-t_2) [H(t_1),H(t_2)],\nonumber\\
K_\text{eff}^{(0)}(t) &=& \bm{0},\nonumber\\
K_\text{eff}^{(1)}(t) &=& \frac{1}{i\hbar\Omega}\sum_{\ell\neq 0}\frac{\mathrm{e}^{i\ell\Omega t}}{\ell} H_\ell = -\frac{1}{2\hbar}\int_{t}^{T+t}\mathrm{d}t'H(t')g(t-t'),
\label{eq:kick_operator_HFE}
\end{eqnarray}
where we Fourier-decomposed the Hamiltonian as $H(t) = \sum_{\ell=-\infty}^\infty H_\ell e^{i\ell\Omega t}$ with operator-valued coefficients $H_\ell$ and the functions $f(x) = (1-2x/T)$ and $g(x) = (1+2x/T)$, $x\in[0,T]$ in the integrands are understood periodic with period $T$~\cite{eckardt_15}. Since we are interested in the low-energy spectrum of the Floquet Hamiltonian, it suffices to calculate $H_\text{eff}$ only. However, we remark that the effective kick operator is crucial for the correct description of the dynamics -- both stroboscopic and non-stroboscopic~\cite{bukov_14_pra,bukov_14}.

\section{Applying the Schrieffer-Wolff Transformation to the driven Fermi-Hubbard Model.}

\begin{figure}[h!]
	\includegraphics[width=0.5\columnwidth]{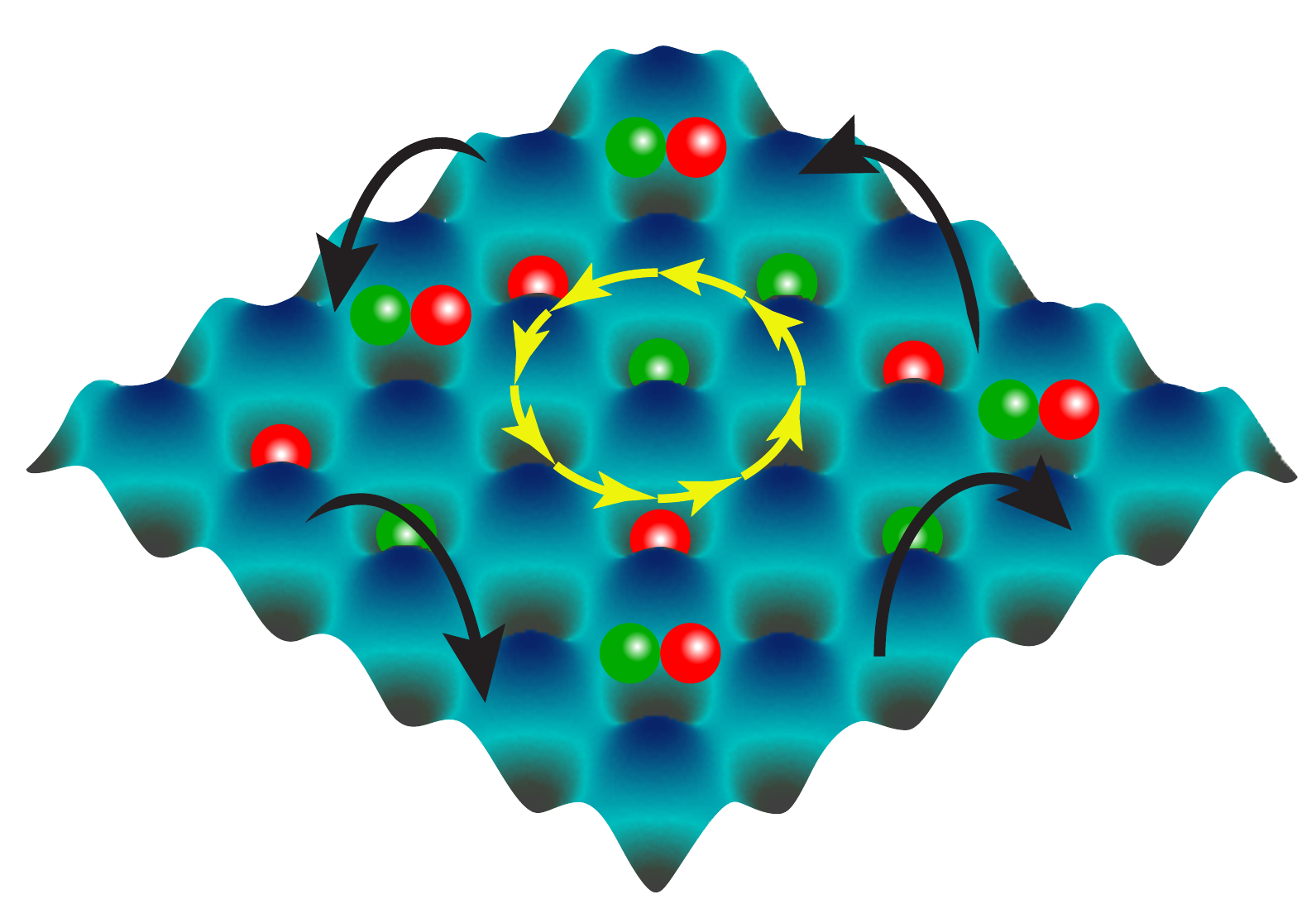}
	\caption{\label{fig:SW_FHM}(Color online). The strongly-interacting Fermi-Hubbard model with an artificial gauge field. }
\end{figure}

In this section, we give the details of the calculation of the effective Hamiltonian in the periodically-driven Fermi-Hubbard model (FHM). The starting point is the Hamiltonian:
\begin{eqnarray}
H(t) = -J_0\sum_{\langle ij\rangle,\sigma}c^\dagger_{i\sigma}c_{j\sigma} + U\sum_j n_{j\uparrow}n_{j\downarrow}+ \sum_{j,\sigma} f_{j\sigma}(t)n_{j\sigma}.
\label{eq:rotframe_H}
\end{eqnarray}
Going to the rotating frame is equivalent to a re-summation of two infinite lab-frame inverse-frequency subseries~\cite{bukov_14}. The first subseries leads to a non-perturbative renormalisation of the hopping amplitude by resumming single-particle terms, while the second subseries contains the many-body nn-interaction-dependent hopping terms. Using the change-of-reference-frame transformation $V(t)=e^{-i U t \sum_j n_{j\uparrow}n_{j\downarrow}}e^{-i\sum_{j,\sigma} F_{j\sigma}(t)n_{j\sigma} }$, we arrive at the Hamiltonian in the rotating frame:
\begin{eqnarray}
H^\text{rot}(t) = -J_0\sum_{\langle ij \rangle,\sigma}\left[ 1 - n_{i\bar{\sigma}}(1-e^{iUt}) \right]e^{i\delta F_{ij,\sigma}(t)}c^\dagger_{i\sigma}c_{j\sigma}\left[ 1 - n_{j\bar{\sigma}}(1-e^{-iUt}) \right],\nonumber
\end{eqnarray}
where $F_{j\sigma}(t)$ is the anti-derivative of $f_{j\sigma}(t)$ and $\delta F_{ij,\sigma} = F_{i\sigma}(t) - F_{j\sigma}(t)$. It is convenient to cast this expression in the following form
\begin{eqnarray}
H^\text{rot}(t) &=& - J_0\sum_{\langle ij\rangle,\sigma} e^{i\delta F_{ij,\sigma}(t) }g_{ij\sigma} 
-J_0\sum_{\langle ij\rangle,\sigma} e^{i\left[\delta F_{ij,\sigma}(t) + Ut\right]}h^\dagger_{ij\sigma} + \text{h.c.},\nonumber\\
h^\dagger_{ij\sigma} &=& n_{i\bar\sigma}c^\dagger_{i\sigma}c_{j\sigma}(1 - n_{j\bar\sigma}),\nonumber\\
g_{ij\sigma} &=& (1 - n_{i\bar\sigma})c^\dagger_{i\sigma}c_{j\sigma}(1 - n_{j\bar\sigma}) +  n_{i\bar\sigma}c^\dagger_{i\sigma}c_{j\sigma}n_{j\bar\sigma}.
\label{eq:FHM_rot_frame}
\end{eqnarray}
The first term in $g_{ij\sigma}$ gives rise to the hopping of holons, while the second one yields hopping of doublons. The term in $h^\dagger_{ij\sigma}$ is, in turn, responsible for creation of doublons and holes. We draw the reader's attention to the fact that the overall sign of the function $\delta F_{ij,\sigma}$ in the Hamiltonian above depends on the direction of hopping. For instance, for a one-dimensional chain with drive $f_{j\sigma}(t) = j\Omega\zeta\cos\Omega t$ the Hamiltonian~\eqref{eq:FHM_rot_frame}, when fully written out, reads
\begin{eqnarray}
H^\text{rot}(t) &=& - J_0\sum_{j,\sigma} e^{i\zeta\sin\Omega t } g_{j+1,j,\sigma} + e^{-i\zeta\sin\Omega t } g_{j,j+1,\sigma} \nonumber\\
&& - J_0\sum_{j,\sigma} e^{iUt}\left( e^{i\zeta\sin\Omega t } h^\dagger_{j+1,j,\sigma} + e^{-i\zeta\sin\Omega t } h^\dagger_{j,j+1,\sigma} \right) + \mathrm{h.c.}
\end{eqnarray} 
Note also that while $g_{j+1,j,\sigma} = g^\dagger_{j,j+1,\sigma}$, $h_{j+1,j,\sigma} \neq h^\dagger_{j,j+1,\sigma}$; in other words destroying a doublon to the left $h_{21,\downarrow}|\cdot,\uparrow\downarrow\rangle=|\uparrow,\downarrow\rangle$ is different from creating a doublon to the left $h^\dagger_{12,\downarrow}|\uparrow,\downarrow\rangle=-|\uparrow\downarrow,\cdot\rangle$.

\emph{Non-driven case.} Let us pause for a moment and check the non-driven case, i.e.~$F_{ij,\sigma} = 0$. Then the terms proportional to $h^\dagger_{ij\sigma}$ vanish in $H_\text{eff}^{(0)}$ after time-averaging over one period $T = 2\pi/U$, cf.~Eq.~\eqref{eq:kick_operator_HFE}. On the other hand, the $g_{ij\sigma}$-terms do not have a time-dependent pre-factor and hence they give rise to the leading-order Hamiltonian. 
\begin{eqnarray}
H_\text{eff}^{(0)}&=&- J_0\sum_{\langle ij\rangle,\sigma} g_{ij\sigma}  = - J_0\sum_{\langle ij\rangle,\sigma} P_{i\bar\sigma} c^\dagger_{i\sigma}c_{j\sigma}   P_{j\bar\sigma},\nonumber\\
P_{i\bar{\sigma}} c^\dagger_{i\sigma}c_{j\sigma}P_{j\bar{\sigma}} &\equiv& 
n_{i\bar\sigma}c^\dagger_{i\sigma}c_{j\sigma}n_{j\bar\sigma}+
(1 - n_{i\bar\sigma})c^\dagger_{i\sigma}c_{j\sigma}(1 - n_{j\bar\sigma}),
\label{eq:Pi}
\end{eqnarray}
where the above expression is understood as the defining relation for the projector $P_{i\sigma}$ which projects out the subspace of doubly-occupied states. The $U^{-1}$-correction as given by Eq.~\eqref{eq:kick_operator_HFE} is proportional to the commutator  $H_\text{eff}^{(1)}\sim J_0^2U^{-1}\sum_{\langle ij\rangle,\sigma}\sum_{\langle kl\rangle,\sigma'}[h_{ij\sigma}^\dagger,h_{kl\sigma'}]$, and results in the familiar Heisenberg spin exchange. Notice that already at this level the calculation for the static model reduces exactly to the standard SW calculation.

\emph{Driven case.} Now let us turn on the periodic drive again. Pay attention how the zeroth order Hamiltonian changes, since the terms proportional to $h^\dagger_{ij\sigma}$, which average to zero in the non-driven case, now remain finite after averaging over one period. These are precisely the doublon association and dissociation processes in the resonant limit $J_0\ll U=l\Omega$ whose physics we discuss in the main text.

\section{ The Strongly-Interacting Periodically-Driven Fermi-Hubbard-Model Away from Half-Filling.  }

In this section, we give the details of the calculations for the resonant and non-resonant driving regimes, for which we derive the low-energy effective Hamiltonian. In the non-resonant case, we use two consecutive SW transformations, applied in the limits $U\ll\Omega$ and $\Omega\ll U$ [these two limits are reconciled in the next section]. We label the lattice sites by $r_j = (m,n)$.

\emph{(i) Non-resonant Driving Limit.} In this regime, we choose spin-dependent periodic driving of the type used to engineer the Harper-Hofstadter Hamiltonian~\cite{aidelsburger_13,miyake_13}:
\begin{equation}
f_{j,\sigma}(t) = \sigma\left[A\cos\left(\Omega t + \phi_{j}\right) + \Omega \vec{e}_x\cdot\vec{r}_j\right],
\label{eq:driving_prot1}
\end{equation} 
where $\sigma$ is the  fermion spin, and $\phi_j = \phi_{mn} = \Phi_{\square}(m+n)$. In this section, we choose $\Phi_\square = \pi/2$, which results in a quarter flux quantum per plaquette. From the definition of the drive, it becomes clear that opposite spin species are subject to opposite gradient potentials. Notice that spin-exchange processes along the $x$-direction are enabled by a resonant absorption of two photons, leading to an effective gauge field for the Heisenberg model at half-filling. We denote by $\zeta = A/\Omega$ the dimensionless interaction strength.

Let us first focus on the regime $J_0\ll\Omega\ll U$ and show the derivation of the effective Hamiltonian comprising the Heisenberg model in an artificial gauge field. We can identify the largest frequency in the problem to be the interaction strength $U$, followed by the driving frequency $\Omega$. Time-scale separation allows us to first perform a SW transformation to the Hamiltonian in Eq.~\eqref{eq:rotframe_H} w.r.t.~the fast period $T_U = 2\pi/U$. In doing so we treat the time-fluctuations in the Hamiltonian due to the driving protocol at frequency $\Omega$ as slow variables, and apply the HFE expansion with the fast period $T_U$ only. This allows us to effectively take the $T_\Omega$-oscillating terms out of the integrals in the HFE, which results in the familiar $t-J$ model in a presence of a $T_\Omega$-periodic drive. The remaining effective  dynamics induced by the drive happens at time-scales $T_\Omega$ and, in the rotating frame, it is governed by the following intermediate Hamiltonian:
\begin{eqnarray}
H^\text{rot}_\text{intermediate}(t) &=& -J_0\sum_{mn,\sigma}P_{m+1,n\bar\sigma} \left( e^{i\delta F_{m+1,n\sigma}(t) }c^\dagger_{m+1,n\sigma}c_{mn\sigma} + \text{h.c.}\right) P_{mn\bar\sigma}\nonumber\\
&& -J_0\sum_{mn,\sigma}P_{m,n+1,\bar\sigma}\left(  e^{i\delta F_{m,n+1,\sigma}(t) }c^\dagger_{m,n+1,\sigma}c_{mn\sigma} + \text{h.c.}\right) P_{mn\bar\sigma}\nonumber\\
&& + \frac{4J_0^2}{U}\sum_{m,n } \left[ S^z_{m+1,n}S^z_{mn} + \frac{1}{2}\left(e^{i2\delta F_{m+1,n\sigma}(t) }S^+_{m+1,n}S^-_{mn} + \text{h.c.}\right) - \frac{n_{m+1,n}n_{mn}}{4}\right] \nonumber\\
&& + \frac{4J_0^2}{U}\sum_{m,n } \left[ S^z_{m,n+1}S^z_{mn} + \frac{1}{2}\left(e^{i2\delta F_{m,n+1,\sigma}(t) }S^+_{m,n+1}S^-_{mn} + \text{h.c.}\right) - \frac{n_{m,n+1}n_{mn}}{4} \right],
\label{eq:XXZ_rot1}
\end{eqnarray}
where, again we drop the holon hopping term to order $J_0^2/U$, as it will be a minor correction to the order-$J_0$ hopping above~\cite{keeling_notes}. If we consider the system away from half-filling, double occupancies are not suppressed and the spin part of the Hamiltonian~\eqref{eq:XXZ_rot1} is merely a correction. The leading effective Hamiltonian away from half-filling after applying the HFE once again with period $T_\Omega$ reads
\begin{eqnarray}
H_\text{eff}^{(0)} &=&-J_0\mathcal{J}_1(2 \zeta_\Phi )\sum_{mn,\sigma}P_{m+1,n,\bar\sigma}\left(  e^{i\phi_{mn}}c^\dagger_{m+1,n,\sigma}c_{mn\sigma} + \text{h.c.}\right) P_{mn\bar\sigma}\nonumber\\
&& -J_0\mathcal{J}_0(2 \zeta_\Phi )\sum_{mn,\sigma}P_{m,n+1,\bar\sigma}\left(  c^\dagger_{m,n+1,\sigma}c_{mn\sigma} + \text{h.c.}\right) P_{mn\bar\sigma}.
\end{eqnarray}
Notice the presence of a gauge field in the hopping of doublons and holons.

We now switch to half filling. Then one can safely neglect the terms in Eq.~\eqref{eq:XXZ_rot1} containing the projectors $P$, as well as the terms proportional to $n_{m+1,n}n_{mn}/4$, similarly to the case for the static SW transformation. Now we apply the HFE again with the slow frequency $\Omega$. Since the leading correction term scales as $J_0^3/(\Omega U)$ we can safely neglect it to obtain
\begin{eqnarray}
H_\text{eff} &\approx& \frac{4J_0^2}{U}\sum_{m,n }  \Big[ S^z_{m+1,n}S^z_{mn} + \frac{\mathcal{J}_2(4 \zeta_\Phi)} {2}\left(e^{2i\phi_{mn} }S^+_{m+1,n}S^-_{mn} + \text{h.c.}\right) +  S^z_{m,n+1}S^z_{mn} + \frac{\mathcal{J}_0(4 \zeta_\Phi)}{2}\left(S^+_{m,n+1}S^-_{mn} + \text{h.c.}\right) \big].\nonumber
\end{eqnarray}
We thus see that in the regime $J_0\ll \Omega\ll U$, applying the SW transformation at half filling leads to the Heisenberg model in an artificial gauge field. We stress that the effective dynamics of the system is best governed by the above effective Hamiltonian for times $t \lesssim  \Omega U/J_0^3$, set by the magnitude of the next-order correction term. Furthermore, choosing $\Omega$ and $U$ to be incommensurate will lead to suppression of resonant effects, thus enhancing the time interval for which time-scale separation holds. This is possible because the spectra of both $H_\text{int}$ and $H_\text{drive}$ are discrete and commensurate.

Let us also briefly discuss the other non-resonant case $J_0\ll U\ll\Omega$. This time the fastest frequency in the problem is the driving frequency $\Omega$, followed by the interaction strength $U$. Thus, we go to the rotating frame w.r.t.~the driving term first:
\begin{eqnarray}
H^\text{rot}_\text{intermediate}(t) &=& -J_0\sum_{mn,\sigma} \left( e^{i\delta F_{m+1,n\sigma}(t) }c^\dagger_{m+1,n\sigma}c_{mn\sigma} + \text{h.c.}\right) \nonumber\\
&& -J_0\sum_{mn,\sigma}\left(  e^{i\delta F_{m,n+1,\sigma}(t) }c^\dagger_{m,n+1,\sigma}c_{mn\sigma} + \text{h.c.}\right)  + U\sum_{mn}n_{mn,\uparrow}n_{mn,\downarrow}.
\end{eqnarray}
Once again we make use of time-scale separation; applying the HFE with period $T_\Omega$ results in the intermediate Hamiltonian to order $\Omega^{0}=1$:
\begin{eqnarray}
H_\text{intermediate}^{(0)} &=& -J_0\mathcal{J}_1\left(2 \zeta_\Phi \right)\sum_{mn,\sigma}\left(  e^{i\phi_{mn}}c^\dagger_{m+1,n,\sigma}c_{mn\sigma} + \text{h.c.}\right)   \nonumber\\
&& -J_0\mathcal{J}_0\left(2 \zeta_\Phi \right)\sum_{mn,\sigma}\left(  c^\dagger_{m,n+1,\sigma}c_{mn\sigma} + \text{h.c.}\right)   \nonumber\\ 
&& +U\sum_{mn}n_{mn,\uparrow}n_{mn,\downarrow}.
\end{eqnarray}
To complete the derivation, all one has to do is to apply the static SW transformation with frequency $U$. This mimics the static SW transformation and directly leads to the following Heisenberg model at any filling
\begin{eqnarray}
H_\text{eff} &\approx& 
-J_0\mathcal{J}_1\left(2 \zeta_\Phi \right)\sum_{mn,\sigma}P_{m+1,n,\bar\sigma}\left(  e^{i\phi_{mn}}c^\dagger_{m+1,n,\sigma}c_{mn\sigma} + \text{h.c.}\right) P_{mn\bar\sigma}  \nonumber\\
&& -J_0\mathcal{J}_0\left(2 \zeta_\Phi \right)\sum_{mn,\sigma}P_{m,n+1,\bar\sigma}\left(  c^\dagger_{m,n+1,\sigma}c_{mn\sigma} + \text{h.c.}\right) P_{mn\bar\sigma} \nonumber\\
&& +J_\text{eff}^{\mathrm{ex},x}\sum_{mn} \Big[ S^z_{m+1,n}S^z_{mn} + \frac{1}{2}\left(e^{2i\phi_{mn}}S^+_{m+1,n}S^-_{mn} + \text{h.c.}\right) - \frac{n_{m+1,n}n_{mn}}{4} \Big] \nonumber\\
&& +J_\text{eff}^{\mathrm{ex},y}\sum_{mn} [ S^z_{m,n+1}S^z_{mn} + \frac{1}{2}\left(S^+_{m,n+1}S^-_{mn} + \text{h.c.}\right) - \frac{n_{m,n+1}n_{mn}}{4} \Big],
\end{eqnarray}
with the effective exchange interactions $J^\text{ex,y}_\text{eff} = 4\left[J_0\mathcal{J}_0\left(2 \zeta_\Phi \right)\right]^2/U$ and $J^\text{ex,x}_\text{eff} = 4\left[J_0\mathcal{J}_1\left( 2 \zeta_\Phi \right)\right]^2/U$. Notice that since $U\ll\Omega$ the leading $\Omega^{-1}$-correction succumbs to the leading $U^{-1}$-Heisenberg model, so our assumption to drop the former is justified.

\emph{(ii) Resonant Driving Limit.} Last, let us focus on the commensurate case $J_0\ll U=\Omega$. Unlike in the main text, we choose the same driving protocol as in Eq.~\eqref{eq:driving_prot1} which allows us to show how to engineer doublon-holon physics in the presence of a gauge field. In this regime, the Hamiltonian $H^\text{rot}$ in Eq.~\eqref{eq:rotframe_H} is indeed periodic with the single frequency $\Omega = U$. Locking the driving frequency to the interaction strength leads to resonances which drastically change the behaviour of the system. Here, we show that they are captured by the HFE, beyond linear response theory. Moreover, this procedure does not suffer from vanishing denominators as is the case in conventional perturbation theory. To this end, we average Eq.~\eqref{eq:rotframe_H} over one period which is equivalent to keeping only the leading order term in the effective Hamiltonian:
\begin{eqnarray}
H^\text{{(0)}}_\text{eff} &=& -J_\text{eff}^x\sum_{mn,\sigma}P_{m+1,n\sigma} \left( e^{i\phi_{mn} }c^\dagger_{m+1,n\sigma}c_{mn\sigma} + \text{h.c.}\right) P_{mn\bar\sigma} -J_\text{eff}^y\sum_{mn,\sigma}P_{m,n+1,\sigma} \left(  c^\dagger_{m,n+1,\sigma}c_{mn\sigma} + \text{h.c.}\right) P_{mn\bar\sigma},\nonumber\\
&& -\sum_{mn,\sigma}\left(K^{L,x}_\text{eff}n_{m,n\bar\sigma}e^{i\phi_{mn}}c^\dagger_{mn\sigma}c_{m+1,n\sigma}(1-n_{m+1,n\bar\sigma}) + 
K^{R,x}_\text{eff}n_{m+1,n\bar\sigma}e^{i\phi_{mn}}c^\dagger_{m+1,n\sigma}c_{mn\sigma}(1 - n_{mn\bar\sigma}) + \text{h.c.} \right) \nonumber\\ 
&&
- K^y_\text{eff}\sum_{mn,\sigma}\left(n_{mn\bar\sigma}c^\dagger_{mn,\sigma}c_{m,n+1\sigma}(1 - n_{m,n+1,\bar\sigma}) -
n_{m,n+1,\bar\sigma}c^\dagger_{m,n+1,\sigma}c_{mn\sigma}(1 - n_{mn\bar\sigma})   +\text{h.c.} \right),
\label{eq:supp_DH}
\end{eqnarray}
with $K^{R,x}_\text{eff} = J_0\mathcal{J}_2(2 \zeta_\Phi)$, $K^{L,x}_\text{eff} = J_0\mathcal{J}_0(2 \zeta_\Phi)$ and $K^y_\text{eff} = J_0\mathcal{J}_1(2 \zeta_\Phi)$. If the resonant periodic drive couples to the interaction strength instead, one can realise homogeneous doublon-holon creation amplitudes along the $x$-direction $K^{L,x}_\text{eff} = K^{R,x}_\text{eff} = J_0\mathcal{J}_2(2 \zeta_\Phi)$, as well as equal-sign doublon-holon amplitudes along the $y$-direction $K^y_\text{eff} = J_0\mathcal{J}_1(2 \zeta_\Phi)$. Note how the resonance condition $U=\Omega$ brings in additional terms in the effective Hamiltonian even in the leading order, which would not be there in the absence of the drive, i.e.~for $A=0$. Hence, these terms are dominant, compared to the Heisenberg model appearing at order $U^{-1}$, and lead to a fundamentally different physics. In fact, they are responsible for enhancing the probability amplitude for doublon association and dissociation processes, in which two particles, initially populating neighbouring sites, are put on top of each other, or vice-versa. The necessary energy $U$ is provided by one driving quantum $\Omega$. We stress that this is a description beyond linear response theory, since the effective Hamiltonian governs the slow dynamics over a multitude of periods, depending on how well the time-scale separation is pronounced.

The presence of double occupancies in strongly-interacting fermions in periodically-modulated optical lattices is intimately related to energy absorption~\cite{kollath_06,huber_09}. It has been shown that the doublon production rate is the same as the energy absorption rate~\cite{tokuno_11,tokuno_12}. The former has been measured in a recent experiment~\cite{greif_11} and a linear increase in time was found for weak driving amplitudes. In general, lattice modulation spectroscopy can be employed to determine the value of the interaction strength in the strongly-interacting limit. Furthermore, the weight of the double occupancy peak contains information about the spin ordering in the system. For example, an anti-ferromagnetic state is more amenable to formation of doublons, compared to a ferromagnetic or a paramagnetic state. Near half-filling, doublon formation has been proposed as a tool to detect an AFM state, expected to appear in the phase diagram of the FHM with repulsive interactions at low temperatures~\cite{sensarma_09}. Previous work studying similar models focused on the weak-driving limit and employed time-dependent perturbation theory to second order [the linear-response term vanishes averaged over one cycle of the drive]~\cite{kollath_06,sensarma_09,huber_09,strohmaier_10}, and Fermi's Golden rule~\cite{hassler_09}. The effective Floquet Hamiltonian in Eq.~\eqref{eq:supp_DH} is clearly non-perturbative and, therefore, allows for an accurate description of the dynamics over multiple cycles of the drive and in the regime of strong amplitudes, $\zeta \gtrsim 1$. For a better precision, one can compute the first leading correction. Micromotion effects can be understood by studying the kick operator.

\section{ Spin Models from the Fermi-Hubbard Model for Generic Off-Resonant Drive.}

In this section, we show how the previous results for off-resonant drive can be derived from the generalised SW transformation described in the main text in greater detail. We show how the two off-resonant limits $U\ll\Omega$ and $\Omega\ll U$ discussed above can be reconciled into one non-resonant regime. In particular, we prove the validity of consecutive application of SW transformations in models with clear time-scale separation, as presented in the previous section.

Consider a generic driving protocol, which gives the rotating frame Hamiltonian in Eq.~\eqref{eq:rotframe_H}:
\begin{eqnarray}
H^\text{rot}(t) &=& - J_0\sum_{\langle ij\rangle,\sigma} e^{i\delta F_{ij,\sigma}(t) }g_{ij\sigma} 
-J_0\sum_{\langle ij\rangle,\sigma} e^{i\left[\delta F_{ij,\sigma}(t) + Ut\right]}h^\dagger_{ij\sigma} + \text{h.c.},\nonumber\\
h^\dagger_{ij\sigma} &=& n_{i\bar\sigma}c^\dagger_{i\sigma}c_{j\sigma}(1 - n_{j\bar\sigma}),\nonumber\\
g_{ij\sigma} &=& (1 - n_{i\bar\sigma})c^\dagger_{i\sigma}c_{j\sigma}(1 - n_{j\bar\sigma}) +  n_{i\bar\sigma}c^\dagger_{i\sigma}c_{j\sigma}n_{j\bar\sigma} ~.
\nonumber
\end{eqnarray}
Since $\delta F_{i j \sigma}$ is $\Omega$-periodic, we can most generally write it in terms of Fourier coefficients:
\begin{equation}
e^{i \delta F_{i j \sigma}(t)} = \sum_\ell A^{(\ell)}_{i j \sigma} e^{i \ell \Omega t} ~.
\label{eq:FT_F_ijs}
\end{equation}
As with the remainder of the paper, we will consider $\Omega = k \Omega_0$ and $U=l \Omega_0$ with $k$ and $l$ relatively prime and $\Omega_0 \gg J_0$. Furthermore, assume that $k, l \gg 1$ such that resonance effects can be ignored and that the state of the system at half-filling has no doublons or holons which, as we have seen, will not be dynamically generated at low orders in the high-frequency expansion. Then the leading correction is of order $1/\Omega_0$ and we will only be interested in the singly-occupied conserving (a.k.a. spin) terms in the expansion.

Before expanding in powers of $1/\Omega_0$, let us quickly comment on properties of the Fourier coefficients $A^{(\ell)}_{i j \sigma}$. While not necessary for all driving protocols, it will be useful in driving spin Hamiltonians to demand that spin up and down are driven oppositely, i.e., $\delta F_{i j \sigma} = -\delta F_{i j \bar \sigma}$. In terms of the Fourier transform, Eq.~\eqref{eq:FT_F_ijs}, this implies that $A^{(\ell)}_{i j \bar \sigma} = (A^{(-\ell)}_{i j \sigma})^\ast$. Similarly, flipping the direction of the bond flips the sign of $\delta F$, so $A^{(\ell)}_{j i \sigma} = (A^{(-\ell)}_{i j \sigma})^\ast$.

The leading correction to the effective Hamiltonian is $H_\mathrm{eff}^{(1)} = \sum_{\ell > 0} [H_\mathrm{rot}^{(\ell)},H_\mathrm{rot}^{(-\ell)}] / \ell \Omega_0$. There are two types of commutators that occur in this sum. The first comes from terms that have no oscillation with frequency $U$, giving commutators of the form:
\begin{equation}
\left[ \sum_{ij\sigma} A^{(\ell)}_{ij\sigma} g_{i j \sigma}, \sum_{i'j'\sigma'} A^{(\ell)}_{i'j'\sigma'} g_{i' j' \sigma'}\right] ~.
\end{equation}
One can readily check that all of these commutators vanish. The second class of commutators are those that are relevant for the Schrieffer-Wolff transformation:
\begin{equation}
\left[ \sum_{ij\sigma} A^{(\ell)}_{ij\sigma} h_{i j \sigma}^\dagger, \sum_{i'j'\sigma'} A^{(-\ell)}_{i'j'\sigma'} h_{j' i' \sigma'} \right] ~.
\end{equation}
These involve terms rotating with $e^{i (U + \ell \Omega) t}$, and thus will be suppressed by a $(U + \ell \Omega)$ denominator. The commutators vanish if $i$, $i'$, $j$, and $j'$ are all different. For $i=i'$ and $j\neq j'$, the non-vanishing commutators correspond to next-neighbor doublon/holon hopping which is suppressed at half filling. Therefore, the only relevant commutators come from $i=i'$ and $j=j'$ or $i=j'$ and $j=i'$. Note that these are the same commutators that were implicitly used in the previous appendices; we explicitly write them out here for clarity. There are four cases.
\begin{itemize}
	\item {\bf $i'=i$, $j'=j$, $\sigma'=\sigma$: } The commutator vanishes trivially.
	\item {\bf $i'=i$, $j'=j$, $\sigma'=\bar \sigma$: } The commutator gives
	\begin{equation}
	\nonumber
	\mathcal C_1 = A_{ij\sigma}^{(\ell)} A_{ij\bar \sigma}^{(-\ell)} c_{i\sigma}^\dagger c_{j \sigma} c_{i \bar \sigma}^\dagger c_{j \bar \sigma} \left[ (1-n_{i\bar \sigma}) n_{j \bar \sigma} (1-n_{i \sigma}) n_{j \sigma} - n_{i \sigma} (1-n_{j\sigma}) n_{i \bar \sigma} (1-n_{j \bar \sigma}) \right]~.
	\end{equation}
	Using properties of $A^{(\ell)}$ discussed above, the coefficient may be rewritten as $|A_{i j \sigma}^{(\ell)}|^2$.
	\item {\bf $i'=j$, $j'=i$, $\sigma'=\sigma$: } The commutator gives 
	\begin{equation}
	\nonumber
	\mathcal C_2 = A_{ij\sigma}^{(\ell)} A_{j i \sigma}^{(-\ell)} (n_{i \sigma} - n_{j \sigma}) n_{i \bar \sigma} (1-n_{j \bar \sigma}) ~.
	\end{equation}
	The coefficient may be rewritten to $|A_{i j \sigma}^{(\ell)}|^2$.
	\item {\bf $i'=j$, $j'=i$, $\sigma'=\bar \sigma$: } The commutator gives
	\begin{equation}
	\nonumber
	\mathcal C_3 = A_{ij\sigma}^{(\ell)} A_{j i \bar \sigma}^{(-\ell)} c_{i\sigma}^\dagger c_{j \sigma} c_{j \bar \sigma}^\dagger c_{i \bar \sigma} \left[ (1-n_{i\bar \sigma}) n_{j \bar \sigma} (1-n_{j \sigma}) n_{i \sigma} - n_{j \sigma} (1-n_{i\sigma}) n_{i \bar \sigma} (1-n_{j \bar \sigma}) \right]~.
	\end{equation}
	The coefficient may be rewritten $A_{i j \sigma}^{(\ell)} A_{i j \sigma}^{(-\ell)}$.
\end{itemize}
For later convenience, we define the above coefficients for $\sigma = \uparrow$ as 
\begin{equation}
\alpha_{ij}^{(\ell)} \equiv A_{i j \uparrow}^{(\ell)} A_{i j \uparrow}^{(-\ell)} ~,~\beta_{ij}^{(\ell)} \equiv |A_{i j \uparrow}^{(\ell)}|^2 ~.
\end{equation}

The first term, $\mathcal C_1$, yields doublon-holon exchange ($|\uparrow \downarrow,0\rangle \leftrightarrow |0,\uparrow \downarrow\rangle$) and is therefore irrelevant at half filling. Up to a constant energy shift, $\mathcal C_2$ and $\mathcal C_3$ correspond to Ising and exchange terms respectively. Thus the effective spin Hamiltonian may be written
\begin{equation}
H_\mathrm{eff}^{(1)}  =  \sum_{\langle ij\rangle,\ell} \frac{J_0^2}{U + \ell \Omega} \left[ \alpha_{ij}^{(\ell)} S_i^+ S_j^- + (\alpha_{ij}^{(\ell)})^\ast S_i^- S_j^+ + 2 \beta_{ij}^{(\ell)} S_i^z S_j^z \right] ~.
\end{equation}
Hence, we see that the general result is an interacting spin-1/2 Hamiltonian where hopping of the spins is accompanied by a phase that depends on properties of the driving. One can now see how to simply take the limits $U \gg \Omega$ and $\Omega \gg U$. First, if $\Omega \gg U$, then only the $\ell=0$ term in the sum survives:
\begin{equation}
H_\mathrm{eff}^{\Omega \gg U}  =  \frac{J_0^2}{U} \sum_{\langle ij\rangle} \left[ \alpha_{ij}^{(0)} S_i^+ S_j^- + (\alpha_{ij}^{(0)})^\ast S_i^- S_j^+ + 2 \beta_{ij}^{(0)} S_i^z S_j^z \right] ~.
\end{equation}
In the opposite limit, $U \gg \Omega$, not only do all the $\ell$'s contribute, but they contribute with equal weight $1/(U + \ell \Omega) \approx 1/U$:
\begin{equation}
H_\mathrm{eff}^{U \gg \Omega} =  \frac{1}{U} \sum_{\langle ij\rangle, \ell} \left[ \alpha_{ij}^{(\ell)} S_i^+ S_j^- + (\alpha_{ij}^{(\ell)})^\ast S_i^- S_j^+ + 2 \beta_{ij}^{(\ell)} S_i^z S_j^z \right] ~.
\end{equation}
This approximation is technically only valid if the sum is dominated by $\ell \ll U / \Omega$. This condition will generally hold because higher $\ell$'s corresponds to higher harmonics of the drive, which have amplitudes $A^{(\ell)}$ that are exponentially suppressed in $\ell$.

Finally, let us apply this formulation to the drive discussed in the main text,
\begin{equation}
f_{m n \sigma} = \sigma [A \cos(\Omega t + \Phi_\square(m+n)) + \Omega m]~.
\label{eq:f_mn_sigma}
\end{equation}
From the second term in Eq.~\eqref{eq:f_mn_sigma} we see that bonds in the $x$-direction and $y$-direction behave differently. In particular, hopping in the positive $x$-direction gives
\begin{equation}
e^{i \delta F_\uparrow^x} \equiv e^{i(F_{m,n,\uparrow} - F_{m+1,n,\uparrow})} = e^{-i \Omega t} e^{i \zeta (\sin(\Omega t + \Phi_\square(m + n)) - \sin(\Omega t + \Phi_\square(m + n + 1)))} = e^{-i \Omega t} e^{i \delta F_\uparrow^y}~.
\end{equation}
Fourier-transforming this simple harmonic driving, one can readily check that
\begin{equation}
A_{y \uparrow}^{(\ell)} = e^{i \ell (\phi_{mn} + (\Phi_\square + \pi)/2)} \mathcal{J}_\ell (2 \zeta_\Phi) ~,
\end{equation}
from which it is clear that $A_x$ is just shifted by one harmonic:
\begin{equation}
A_{x \uparrow}^{(\ell)} = e^{i (\ell+1) (\phi_{mn} + (\Phi_\square + \pi)/2)} \mathcal{J}_{\ell+1} (2 \zeta_\Phi) ~.
\end{equation}
This gives coefficients on the spin Hamiltonian of
\begin{eqnarray}
\nonumber \alpha_y^{(\ell)} &=& A_y^{(\ell)} A_y^{(-\ell)} = \mathcal{J}_{\ell}(2 \zeta_\Phi) \mathcal{J}_{-\ell}(2 \zeta_\Phi) 
\\ \nonumber \alpha_x^{(\ell)} &=& e^{2 i (\phi_{mn} + (\Phi_\square + \pi)/2)} \mathcal{J}_{\ell+1}(2 \zeta_\Phi) \mathcal{J}_{-\ell + 1}(2 \zeta_\Phi) 
\\ \nonumber \beta_y^{(\ell)} &=& \left[\mathcal{J}_{\ell}(2 \zeta_\Phi) \right]^2
\\ \beta_x^{(\ell)} &=& \left[\mathcal{J}_{\ell+1}(2 \zeta_\Phi) \right]^2 ~.
\end{eqnarray}
The overall phase factor $\Phi_\square + \pi$ in $\alpha_x$ is irrelevant to the global physics, so we gauge it away by rotating $S^+_{m n} \to S^+_{mn} e^{i m (\Phi_\square + \pi)}$. Then, for $\Omega \gg U$ the Hamiltonian reduces to 
\begin{eqnarray}
\nonumber H_\mathrm{eff}^{\Omega \gg U}  &  = &  \frac{2 J_0^2}{U} \sum_{mn} \Big[ [\mathcal{J}_1(2 \zeta_\Phi)]^2 ( e^{2 i \phi_{mn}} S_{m,n}^+ S_{m+1,n}^- + e^{-2 i \phi_{mn}} S_{m,n}^- S_{m+1,n}^+ + 2 S_{m,n}^z S_{m+1,n}^z)
\\ && ~~~~~~~~ + [\mathcal{J}_0(2 \zeta_\Phi)]^2 ( S_{m,n}^+ S_{m,n+1}^- + S_{m,n}^- S_{m,n+1}^+ + 2 S_{m,n}^z S_{m,n+1}^z) \Big] ~.
\label{eq:supp_U<<Omega}
\end{eqnarray}
The $U \gg \Omega$ limit can be obtained by using sum rules for the Bessel functions: $\sum_\ell \alpha_x^{(\ell)} = \mathcal{J}_2(4 \zeta_\Phi)$, $\sum_\ell \alpha_y^{(\ell)} = \mathcal{J}_0(4 \zeta_\Phi)$, and $\sum_\ell \beta_{x/y}^{(\ell)} = 1$. Thus,
\begin{eqnarray}
\nonumber H_\mathrm{eff}^{U \gg \Omega}  & = &  \frac{2 J_0^2}{U} \sum_{mn} \Big[ \mathcal{J}_2(4 \zeta_\Phi) ( e^{2 i \phi_{mn}} S_{m,n}^+ S_{m+1,n}^- + e^{-2 i \phi_{mn}} S_{m,n}^- S_{m+1,n}^+) 
\\ && ~~~~~~~~ + \mathcal{J}_0(4 \zeta_\Phi) ( S_{m,n}^+ S_{m,n+1}^- + S_{m,n}^- S_{m,n+1}^+) + 2 S_{m,n}^z S_{m+1,n}^z + 2 S_{m,n}^z S_{m,n+1}^z \Big] ~.
\label{eq:supp_Omega<<U}
\end{eqnarray}

\section{ Reconciling the Resonant and Off-Resonant Limits. Crossover Regime. }

Since the argument we used in the main text for the Floquet realisation of strongly-correlated condensed matter models relies on a clear time-scale separation, it is interesting to explore how the three limits of (i) high-frequency $J_0\ll U\ll \Omega$, (ii) strong interactions $J_0\ll \Omega \ll U$, and (iii) resonant driving $J_0\ll U=l\Omega$ can be reconciled to reproduce the stroboscopic dynamics of the system in the presence of the drive. To illustrate this, it suffices to consider the driven two-site Hubbard model. Thus, we also leave aside the gauge fields which would only obscure the equations. The Hamiltonian is
\begin{eqnarray}
H(t) = -J_0\sum_{\sigma}\left(c^\dagger_{1\sigma}c_{2\sigma} + \text{h.c.}\right) + A\cos(\Omega t)n_2 + U(n_{1\uparrow}n_{1\downarrow} + n_{2\uparrow}n_{2\downarrow} ).
\end{eqnarray}  
Following the discussion and notation of Eq.~\eqref{eq:FHM_rot_frame}, we find the following rotating-frame Hamiltonian
\begin{eqnarray}
H^\text{rot}(t) &=& - J_0\sum_{\sigma}\left( \gamma^*(t) g_{12\sigma} + \text{h.c.} \right)
-J_0\sum_{\sigma}\left( \chi_R^*(t)h^\dagger_{21\sigma} + \chi_L^*(t)h^\dagger_{12\sigma} + \text{h.c.} \right),\nonumber\\
\gamma^*(t) &=& e^{i \zeta\sin\Omega t }, \ \ \ \ \chi_R^*(t) = e^{i(\zeta\sin\Omega t + Ut)}, \ \ \ \ \chi_L^*(t) = e^{-i(\zeta\sin\Omega t - Ut)},
\end{eqnarray}
where the operators $h^\dagger_{ij\sigma}$ and $g_{ij\sigma}$ are defined in Eq.~\eqref{eq:FHM_rot_frame}, and $\zeta = A/\Omega$. 

As already mentioned in the main text, in general the Hamiltonian $H^\text{rot}(t)$ is neither periodic with the frequency $\Omega$, nor with the frequency $U$. In order to apply the high-frequency expansion, we first find two co-prime integers $l$ and $k$ such that $\Omega = k\Omega_0$ and $U = l\Omega_0$, where $\Omega_0 = 2\pi/T_{\Omega_0}$ is the common frequency, such that $H(t+T_{\Omega_0}) = H(t)$. We first need to Fourier-expand the functions $\gamma^*(t)$ and $\chi^*(t)$ in this common frequency $\Omega_0$. Note that, in principle, in order to apply the HFE, one needs to make sure that $J_0\ll \Omega_0$ which may not be true. However, as we shall see shortly, this condition is somewhat artificial since $\Omega_0$ is not a physical scale but rather a mathematical construct. From the Jacobi-Anger identity it follows that $\chi_R^*(t) = \sum_{\ell=-\infty}^\infty \mathcal{J}_\ell(\zeta)e^{i(\ell k+l)\Omega_0 t}\equiv\sum_{\ell=-\infty}^\infty a^R_\ell e^{i\ell\Omega_0 t}$ and similarly for $\chi_L^*(t)$. Clearly, $\chi_R^*(t)$ has a non-zero time average. On the other hand, one can convince oneself that the coefficient $a_0$ is nonzero if and only if $l=-\ell k$. However, since $l$ and $k$ are co-prime, this can only hold true for $k=1$ which means $U=l\Omega$. Physically, this condition is a manifestation of the conservation of quasienergy, saying that the doublon-holon creation term $h^\dagger_{ij\sigma}$ is non-zero at the level of the time-average Hamiltonian only when the interaction strength matches a multiple of the driving frequency.  

We therefore focus only on the resonant case $U=l\Omega$, for which we find $a^R_\ell = \mathcal{J}_{\ell-l}(\zeta)$ and $a^L_\ell = \mathcal{J}_{-\ell-l}(\zeta)$. Fourier-decomposing the Hamiltonian in the rotating frame immediately leads to
\begin{eqnarray}
H_\ell = - J_0\sum_{\sigma} \Big[ \mathcal{J}_{\ell}(\zeta)g_{12\sigma} + \mathcal{J}_{-\ell}(\zeta)g^\dagger_{12\sigma} + \mathcal{J}_{\ell-l}(\zeta)h^\dagger_{21\sigma} + \mathcal{J}_{-\ell+l}(\zeta)h^\dagger_{12\sigma} + \mathcal{J}_{-\ell-l}(\zeta)h_{21\sigma} + \mathcal{J}_{\ell+l}(\zeta)h_{12\sigma} \Big]. 
\end{eqnarray}
Following Eq.~\eqref{eq:kick_operator_HFE}, all leading correction terms can be obtained from $H^{(1)}_\text{eff} = \sum_{\ell\neq 0}[H_\ell,H_{-\ell}]/\ell\Omega_0$. For simplicity let us concentrate on the spin exchange term only, which is proportional to the commutator $[h^\dagger_{ij\sigma},h_{ji\bar\sigma}]$~\cite{keeling_notes}. One can shift the index of the Bessel functions in the sum over $\ell$, and after some algebra we obtain the resonant drive-renormalised exchange interaction $J^\text{ex}_\text{eff}$ as
\begin{eqnarray}
J^\text{ex}_\text{eff} = 4\frac{J_0^2}{\Omega_0}\sum_{\substack{\ell=-\infty \\ \ell\neq-l }}^{\infty} \frac{\mathcal{J}^2_\ell(\zeta)}{l+\ell} = 4\frac{J_0^2}{U}\sum_{\substack{\ell=-\infty \\ \ell\neq-l }}^{\infty} \frac{\mathcal{J}^2_\ell(\zeta)}{1+\ell/l} = 4\frac{J_0^2}{U}\sum_{\substack{\ell=-\infty \\ \ell\neq-l }}^{\infty} \frac{\mathcal{J}^2_\ell(\zeta)}{1+\ell\Omega/U},
\label{eq:exchange_renormalised}
\end{eqnarray}
where in the second and third equalities we used $U = l\Omega_0 = l\Omega$ on resonance. We can now analytically continue $U/\Omega$ from an integer to the entire real axis. In doing so, note that the restriction in the summation $\ell\neq -l$ is superfluous for all non-integer values of $U/\Omega$, i.e.~everywhere away from resonance. This expression was first derived in Ref.~\cite{mentink_15} using an extended Hilbert space approach, which is different but equivalent~\cite{eckardt_15} to the one presented in our work. We note in passing that the renormalisation of the spin-exchange coupling is the same, no matter whether the periodic driving couples to the density (as in our case) or to the interaction strength. The general validity of this type of analytic continuation is a subject of current investigation. It is clear that it will fail for nearly-resonant drives, but these cases can be treated introducing a small detuning $\delta$ to separate out the resonant part already in the lab frame. Nevertheless, we have verified that this procedure produces the correct answer also in the derivation of the Kondo model from the Anderson model where the two incommensurate energy scales are given by the interaction strength $U$ for two electrons occupying the impurity level, and the relative shift $V$ of the impurity level w.r.t.~the Fermi sea.   

Let us now briefly discuss the three limits of interest from the point of view of the general expression, Eq.~\eqref{eq:exchange_renormalised}. Consider first $J_0\ll \Omega\ll U$. In this case, we can safely drop the restriction on the summation and, using the `trigonometric' identity $\sum_\ell \mathcal{J}^2_\ell(\zeta) = 1$, we find the same exchange interaction as in the non-driven model, $J^\text{ex}_\text{eff} = 4J_0^2/U$. This is consistent with first doing the SW transformation w.r.t~$U$ and then applying the FHE w.r.t~$\Omega$, as explained in the main text. The Bessel functions which appear in front of the $S^+S^-$ terms in Eqs.~\eqref{eq:supp_U<<Omega} and~\eqref{eq:supp_Omega<<U} are due to the spin-dependent drive and are not present for spin-independent protocols as the one considered in this section. In the high-frequency regime $J_0\ll U\ll \Omega$, only the $\ell=0$ term contributes, and we find $J^\text{ex}_\text{eff} = 4J_\text{eff}^2/U$, with $J_\text{eff} = J_0\mathcal{J}_0(\zeta)$. Again, this is exactly what one would expect from first applying the HFE to obtain the FHM with renormalised hopping amplitude, and subsequently doing the SW transformation (see main text). Last, the resonant case $J_0\ll U=l\Omega$ is clear from the derivation above. Note, however, that the exchange physics is of order $1/\Omega$, and hence it succumbs entirely to the doublon-holon physics in this regime.

\section{ Loading Sequence for the Ground State of the Effective Doublon-Holon model on Resonance. }

Let us briefly comment on a possible procedure to load the system in the ground state of the effective Floquet Hamiltonian $H_\mathrm{eff}^{(0)}$, describing the doublon-holon model for resonant driving:
\begin{eqnarray}
H_\text{eff}^{(0)} =  \sum_{\langle ij\rangle,\sigma} \left[  - J_\text{eff} g_{ij\sigma}  - K_\text{eff}\!\left( (-1)^{l\eta_{ij}} h^\dagger_{ij\sigma} + \text{h.c.}\right)\right],
\end{eqnarray} 
where $\eta_{ij} = 1$ for $i>j$, $\eta_{ij} = 0$ for $i<j$, $J_\text{eff} = J_0\mathcal{J}_0(\zeta)$, $K_\text{eff} = J_0\mathcal{J}_l(\zeta)$, and $U=l\Omega$. In the following, we concentrate on the case $l=2$ which, as we have shown in the main text, contains a free-fermion point for $J_\mathrm{eff} = K_\mathrm{eff}$.

While most experimental realisations of Floquet Hamiltonians use an adiabatic ramp up of the driving protocol by gradually switching on the drive amplitude~\cite{pweinberg_15}, we follow a slightly different approach, which we find to be more efficient in this case. To minimise heating effects due to resonant absorption from the drive, we use a multi-step loading procedure similar to the one proposed in Ref.~\cite{tsuji_11}. First, at time $t=0$ we prepare the system in the ground state of free fermions. Then we suddenly quench-start the drive [including the interactions] with an amplitude corresponding to the free fermion point: $\zeta = A/\Omega \approx 1.8412$. This procedure preserves the state to leading order in the HFE. Second, we ramp down the driving amplitude smoothly into the Luttinger liquid phase for a total of forty driving periods and stop whenever the amplitude reaches a value such that $K_\mathrm{eff}/(K_\mathrm{eff}+J_\mathrm{eff})\approx 0.2$. Last, we evolve the system at this constant final amplitude for five more driving periods. We note in passing that a similar procedure works when the amplitude is instead increased and the system enters the gapped bond density wave phase.

\begin{figure}[!htb]
	\centerline{%
		\includegraphics[width=0.45\textwidth]{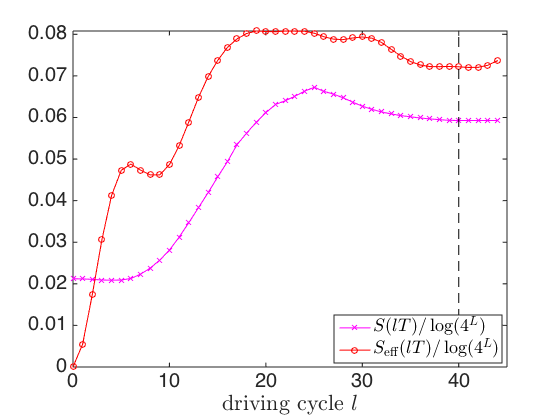}%
	}		
	\caption{ Stroboscopic time evolution of the diagonal entropies for the ramp into the Luttinger Liquid phase on resonance for a chain of $L=8$ sites. The dashed vertical line marks the end point of the ramp, after which the evolution continues at a constant driving amplitude. Unity on the $y$-axis corresponds to maximum entropy while zero -- to minimum. The parameters are $U/J_0=40, U=2\Omega$, $A_i/\Omega = 1.8412\ [J_\mathrm{eff}=K_\mathrm{eff}]$ and $A_f/\Omega = 1.2\ [K_\mathrm{eff}/(J_\mathrm{eff}+K_\mathrm{eff})\approx 0.2$].}
	\label{fig:ramp_entropy}
\end{figure}

To measure the amount of non-adiabaticity introduced during the ramp process, we compute numerically the diagonal entropy in the Floquet eigenbasis, which effectively measures occupation of higher-energy Floquet states. Let us denote by $|\psi(t)\rangle$ the state, exactly evolved with the full lab-frame time-dependent Hamiltonian $H(t)$, whose driving amplitude is ramped down smoothly. Further, we denote the set of eigenstates of the leading-order Floquet Hamiltonian $H_\mathrm{eff}^{(0)}$ by $\{|\nu\rangle\}$, and the probability to be in each of these states at the stroboscopic times $t=lT$ is given by $p^\mathrm{eff}_{\nu\psi}(lT) = |\langle\psi(lT)|\nu\rangle|^2$. While calculating the fidelity requires the unique identification of the Floquet ground state [more precisely the adiabatically-connected Floquet state] at each point of time, we choose to look at the stroboscopic Floquet diagonal entropy $S_\mathrm{eff}(lT) = -\sum_{\nu}p^\mathrm{eff}_{\nu\psi}(lT)\log p^\mathrm{eff}_{\nu\psi}(lT)$, which measures the spread of the initial state over the  basis of the approximate Floquet Hamiltonian as a function of time~\cite{pweinberg_15}. A small value of the entropy automatically means that the system predominantly occupies a single state without the need of identifying it. 

Since the Hamiltonian $H_\mathrm{eff}^{(0)}$ is just the zeroth order term in the HFE, and because any realistic experimental set-up requires a finite frequency, it is also interesting to study the effect of the higher-order terms. This can be done along the same lines by defining the exact Floquet states $\{|n\rangle \}$, and the corresponding probabilities $p_{n\psi}(lT) = |\langle\psi(lT)|n\rangle|^2$ and diagonal entropy $S(lT) = -\sum_{n}p_{n\psi}(lT)\log p_{n\psi}(lT)$. The entropy $S_\mathrm{eff}$ shows how close the state is to the desired ground state of the Hamiltonian $H_\mathrm{eff}^{(0)}$, while the entropy $S$ shows how close the state is to the ground state of the exact instantaneous, i.e.~stroboscopic, Floquet Hamiltonian $H_F$, which knows about the higher-order correction terms.

Figure~\ref{fig:ramp_entropy} shows the two entropies during the ramp. Notice that the nonadiabatic (and hence heating) rates are minimal. This plot also implies that the exact Floquet ground state is very close to the ground state of the approximate Floquet Hamiltonian $H_{\rm eff}^0$. We have verified that a longer ramp duration corresponds to smaller heating rates. We also checked that the mismatch between the two entropies decreases with increasing the drive frequency, according to expectations. While we cannot numerically verify the feasibility of such a loading scheme for larger systems, based on the DMFT results of Ref.~\cite{tsuji_11} where a similar procedure has been employed, we believe that this protocol should be robust even in thermodynamic limit, as the heating effects due to the drive are at most exponentially slow in frequency~\cite{kuwahara_15,abanin_15,abanin_15_2,mori_15} and should not play any role during the finite-time loading process. Therefore, we anticipate that such a protocol will allow one to load larger systems into a low-entropy state which is close enough to the desired Floquet ground state in order to detect the corresponding Luttinger liquid physics.

\end{widetext}

\end{document}